\documentclass[twocolumn,superscriptaddress,preprintnumbers,amsmath,amssymb,floatfix,nofootinbib]{revtex4-1}

\usepackage{graphicx}
\usepackage{multirow}
	
\begin{document}

\title{Noise-Induced Schooling of Fish}

\author{Jitesh Jhawar}
\thanks{These authors contributed equally to the manuscript.}
\affiliation{Centre for Ecological Sciences, Indian Institute of Science, Bangalore, India.}

\author{Richard G.~Morris}
\thanks{These authors contributed equally to the manuscript.}
\affiliation{Simons Centre for the Study of Living Machines, National Centre for Biological Sciences, TIFR, Bangalore, India.}
\affiliation{EMBL-Australia, University of New South Wales, Sydney, Australia.}

\author{U.~R.~Amith-Kumar}
\affiliation{Centre for Ecological Sciences, Indian Institute of Science, Bangalore, India.}

\author{M.~Danny~Raj}
\affiliation{Department of Chemical Engineering, Indian Institute of Science, Bangalore, India.}

\author{Harikrishnan R.}
\affiliation{Centre for Ecological Sciences, Indian Institute of Science, Bangalore, India.}

\author{Vishwesha Guttal}
\affiliation{Centre for Ecological Sciences, Indian Institute of Science, Bangalore, India.}

\begin{abstract}
		We report on the dynamics of collective alignment in groups of the cichlid fish, {\it Etroplus suratensis}. Focusing on small-to-intermediate sized groups ($10\lesssim N\lesssim 100$), we demonstrate that schooling (highly polarised and coherent motion) is noise-induced, arising from the intrinsic stochasticity associated with finite numbers of interacting fish.  The fewer the fish, the greater the (multiplicative) noise and therefore the likelihood of alignment.  Such empirical evidence is rare, and tightly constrains the possible underlying interactions between fish: computer simulations indicate that {\it E. suratensis} align with each other one at a time, which is
		at odds with the canonical mechanism of collective alignment, local direction-averaging. More broadly, our results confirm that, rather than simply obscuring otherwise deterministic dynamics, noise is fundamental to the characterisation of emergent collective behaviours, suggesting a need to re-appraise aspects of both collective motion and behavioural inference.
\end{abstract}

\maketitle

Over the past decade, modern methods of image analysis and tracking have been used extensively to study the collective motion of animal groups, and by proxy the social interactions between their constituent individuals.
However, whilst both techniques and taxa have varied, spanning flocks of starlings \cite{Ballerini2008,Cavagna2010,Bialek2012a,Pearce2014a,Attanasi2014}, shoals of fish \cite{Becco2006,Katz2011a,Herbert-Read2011,Gautrais2012,Ward2017a,Jiang2017,Calovi2018}, marching locusts \cite{Buhl2006,Yates2009a}, mice \cite{Shemesh2013} and red deer \cite{Rands2014}, such empirical studies almost exclusively overlook the {\it intrinsic noise} that arises in any collective, or group  whose underlying individuals interact in an inherently probabilistic way. Only in \cite{Dyson2015}--- a theoretical follow-up to the pioneering study of direction-switching in locust nymphs \cite{Yates2009a}--- are such ideas used in light of experimental evidence, albeit with different conclusions to those described here (see {\it Discussion}).

This is a significant oversight; system-size expansions of Master equations \cite{van_kampen,Gar03} readily demonstrate that probabilistic individual behaviours can conspire to produce collective features that are wholly surprising.  Such approaches are typically referred to as {\it mesoscopic} since they describe the properties of large-but-finite sized groups, where the stochastic behaviour of the individuals cannot be completely `averaged-out'. Notably, this residual stochasticity often manifests as a multiplicative, or state-dependent noise at the collective level, the ramifications of which are not otherwise captured by either its deterministic ($N\to\infty$) or {\it ad hoc} additive-noise counterparts (the latter being typical of a broad class of hydrodynamic descriptions \cite{Hohenberg1977,Toner1995}).  For instance, finite-sized groups of individuals who either copy each other at random or spontaneously change their mind, are predicted to arrive at a clear consensus when faced with a binary choice (such as food sources, for example), whereas a deterministic description predicts no such agreement \cite{TBLDAJM14}.

Such `noise-induced' character \cite{Horsethemke2006,Ridolfi2011} is particularly relevant when trying to {\it infer} individual behaviours from that of the collective.  Here, rather than simply obscuring the signature of otherwise deterministic dynamics, collective-level noise actually encodes important information about individual interactions \cite{Boettiger2018}.  Therefore, not only should fluctuations be extracted with care, but they are also pertinent to a major challenge of behavioural inference: how to distinguish between multiple mechanisms that ostensibly reproduce the same qualitative features of collective motion.

It is in this context that we report on the statistics of directional alignment in freshwater fish ({\it Etroplus suratensis}) under controlled laboratory conditions. We use a data-driven approach, formally extracting a stochastic differential equation (SDE) that describes the dynamics of collective alignment. We find that schooling in such fish--- {\it i.e.}, highly polarised and coherent motion--- bears all the hallmarks of a noise-induced effect, resulting from finite sized groups of individuals that interact according to probabilistic rules. Put simply, the smaller the number of fish in a group, the larger the stochastic fluctuations and surprisingly, the greater the ordering.  This counter-intuitive result can be traced-back to an $O(1/\sqrt{N})$ noise term that is multiplicative--- {\it i.e.}, where the strength of noise depends on the collective state of the group.

\begin{figure*}[t]
	\centering
	\includegraphics[width=0.98\textwidth]{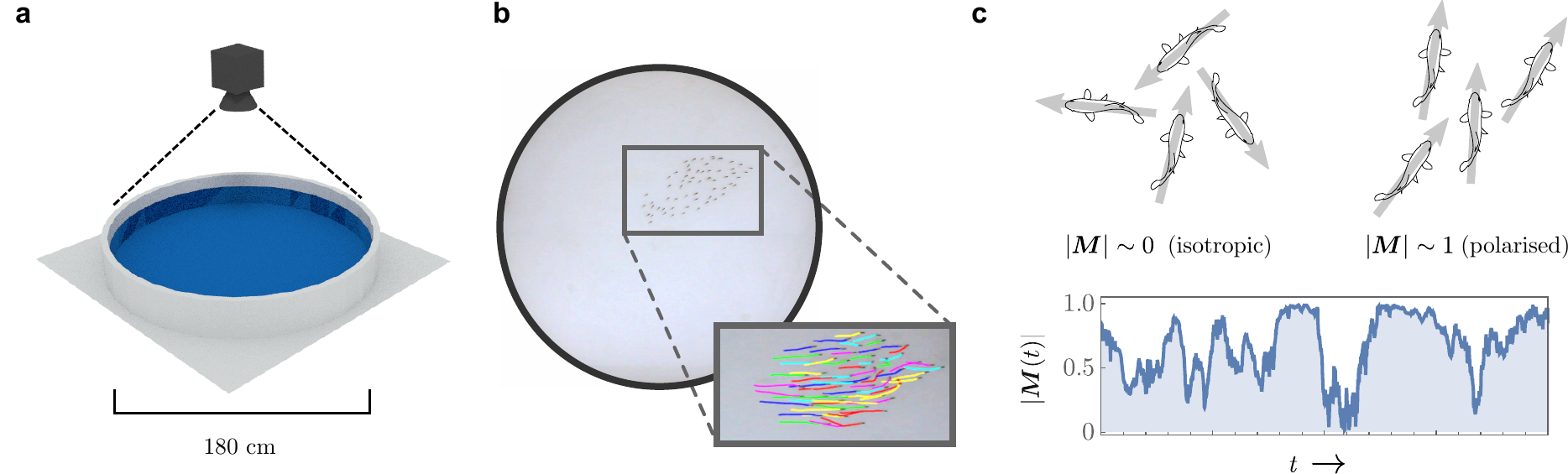}
	\caption
	{
		(Colour online) {\bf Capturing the statistics of ordering in fish}.  Schools of juvenile {\it Etroplus suratensis} were filmed in a large shallow tank (180 cm diameter, 10 cm height), under controlled laboratory conditions (panel {\bf a}).  Using particle tracking methods, we obtained two-dimensional trajectories for each individual fish (panel {\bf b}).  To quantify the coherence of motion, we constructed the group polarisation $\boldsymbol{M}(t)$ [see Eq.~(\ref{eq:m})] such that, for $\left\vert \boldsymbol{M}\right\vert\approx 1$ the fish are moving in a coherent direction, whereas for $\left\vert \boldsymbol{M}\right\vert\approx 0$, there is no prevailing direction and the shoal is effectively isotropic (panel {\bf c}).}
	\label{fig:1}
\end{figure*}
Significantly, the type of finite-size noise-induced schooling that we observe tightly constrains the possible interactions between fish that might underpin it.  Using computer simulations, we find that all the salient features are captured by a simple stochastic protocol for pairwise interactions, whereby a given fish interacts and aligns with other (single) fish, one at a time.  Moreover, such features are {\it not} present if ternary, or higher-order aligning interactions are dominant, including local directional averaging, as used in the Vicsek-like family of approaches \cite{Vicsek1995,Toner1995,Czirok1999}--- the {\it de facto} standard for modelling collective motion.

\section{Experimental setup}
Our experiments involved filming schools of freshwater juvenile {\it E. suratensis} in a large laboratory tank that was sufficiently shallow to effectively constrain motion to two dimensions (see Fig.~\ref{fig:1}).  We focussed on small-to-medium sized groups--- $N=15$, $N=30$, and $N=60$--- which ensured that schools were well-defined, with a spatial extent that was only a fraction of the overall tank size [see Secs.~III and IV of the Supplementary Material (SM) for details concerning both larger schools and controls that rule out significant boundary effects].

Each group size was recorded for approximately 3.5 hours in total, across four separate trials for $N=15$, $30$, and three separate trials for $N=60$ (see Sec.~II, SM).  The temporal resolution was one frame every $0.04\,\mathrm{s}$.  However, at this time-scale, movement is intermittent and the fish frequently stop and start, making it hard to discern a direction of motion.  We therefore consider only one frame in every three--- that is, every $\delta t= 0.12\,\mathrm{s}$.  Using particle tracking (Sec.~II, SM), we extract the two-dimensional velocities $\boldsymbol{v}_i(t_n) =\left[\boldsymbol{x}_i(t_n+\delta t)-\boldsymbol{x}_i(t_n)\right]/\delta t$, where the index $i=1,2,\ldots,N$ labels fish, and the time increments $t_n = n\,\delta t$, for $n=0,1,2,\ldots$ {\it etc}.

At the individual level, the direction of motion of the $i$-th fish (at time $t_n$) is just $\hat{\boldsymbol{v}}_i(t_n) = \boldsymbol{v}_i(t_n)/\left\vert\boldsymbol{v}_i(t_n)\right\vert$.  At the group level, both the direction and degree of the fish alignment are encapsulated by a vector order parameter, also referred to as group polarisation
\begin{equation}
	\boldsymbol{M}(t_n) = \frac{1}{N}\sum_{i=1}^N 
	\hat{\boldsymbol{v}}_i(t_n).
	\label{eq:m}
\end{equation}
When $\left\vert\boldsymbol{M}\right\vert$ is close to 1, the fish are moving in a coherent direction, whereas when $\left\vert\boldsymbol{M}\right\vert$ is close to zero, there is no prevailing direction and individual motion is effectively isotropic (see Fig.~\ref{fig:1}).  

\section{Schooling increases as group size decreases}
\begin{figure*}[t]
	\centering
	\includegraphics[width=0.98\textwidth]{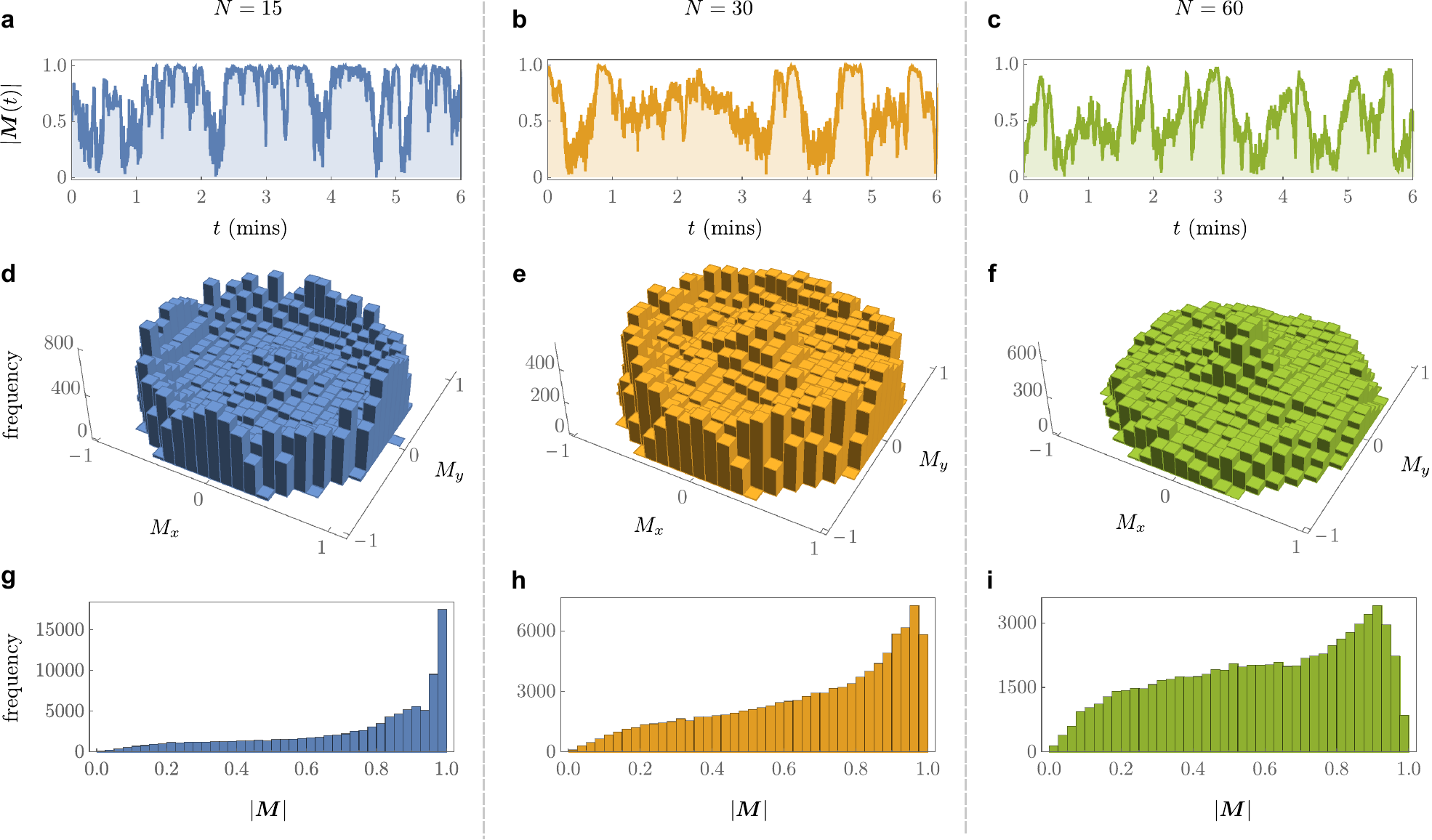}
	\caption
	{
		(Colour online) {\bf Steady-state statistics of ordering}.  A visual inspection of representative trajectories $\vert\boldsymbol{M}(t)\vert$ indicates that, as group size increases, fish motion becomes less ordered (panels {\bf a}-{\bf c}).  Histograms of $\boldsymbol{M}$ reveal there is no preferred direction of motion (panels {\bf d}-{\bf f}).  Taking account of this angular symmetry, panels {\bf g}-{\bf i} demonstrate that highly ordered motion is the dominant behaviour, and confirm the size-dependent nature of ordered motion.  Results are independent of the time-interval over which the histograms are constructed, so long as it is larger than the correlation time.}
	\label{fig:2}
\end{figure*}
At each group size, the steady-state statistics of the school's polarisation are well represented by constructing histograms over the entire time-series (see Sec.~II of SM and Fig.~\ref{fig:2}, panels {\bf d}-{\bf f}).

For $N=60$, the most likely configuration is around $\boldsymbol{M}=\boldsymbol{0}$, which corresponds to motion that is isotropic and disordered (Fig.~\ref{fig:2}, panel {\bf d}).  For $N=30$, this isotropic peak is still present in the histogram, but reduced, and an annulus of high frequencies can be seen where $\left\vert \boldsymbol{M}\right\vert\approx 1$, corresponding to highly aligned motion (Fig.~\ref{fig:2}, panel {\bf e}).  For $N=15$, the highest frequency values are near the highly-aligned $\left\vert \boldsymbol{M}\right\vert\approx 1$ annulus, with only a small isotropic bump at the centre (Fig.~\ref{fig:2}, panel {\bf f}).

Regardless of group size, the statistics have angular symmetry, which suggests that there is no preferred direction of schooling, as might be expected. Considering only the magnitude of the polarisation (see Fig.~\ref{fig:2}, panels {\bf g}-{\bf i}), the likelihood of observing isotropic motion relative to that of ordered motion is revealed to be almost negligible, despite the central peaks observed in panels {\bf d}-{\bf f}.  More importantly, the relative likelihood of observing fish with highly aligned motion {\it increases} as $N$ {\it decreases}, which is also supported by visual inspection of the underlying trajectories (Fig.~\ref{fig:2}, panels {\bf a}-{\bf c}).

\section{Schooling is a noise-induced effect}
To understand the dynamical context of the our observations, we extract autocorrelation functions for the group polarisation components $M_x$ and $M_y$ (Sec.~V, SM).  The results are qualitatively similar to exponentially-damped cosines, indicating the presence of two characteristic time-scales: one for the short-time decay of correlations to zero, and the other for the decaying envelope of longer-time oscillatory behaviour.  The latter appears to be consistent with the effects of finite tank size--- on average, the speed of the fish is $\approx 6\,$cm$\,$s$^{-1}$ which, given a tank diameter of $180\,$cm, implies a time-scale of approximately $30\,$s, in-line with the observed range of 20-50$\,$s.  We therefore focus on the shorter time-scale, whose mean value (across all experiments) is $\Delta t = 49\,\delta t\approx 5.9\,\mathrm{s}$. This, we assert, captures correlations (and their decay) due to any underlying local interactions that result in the alignment of fish.

On the time-scale $\Delta t$, we further assume (and later confirm) that the dynamics of alignment is well-approximated by a stochastic differential equation (SDE), of the type that arises from system-size expansions of transition rates \cite{van_kampen,Gar03} and is synonymous with steady-state statistics that are $N$-dependent.  Usually, such mesoscopic SDEs are constructed formally by coarse-graining known `microscopic' rules.  However, in the field of collective behaviour these are, of course, unknown, describing the individuals and their behavioural interactions.  We therefore quantify the mesoscopic dynamics directly from the data, and then later infer microscopic rules by comparison with computer simulations.

\begin{figure*}[!t]
	\centering
	\includegraphics[width=0.98\textwidth]{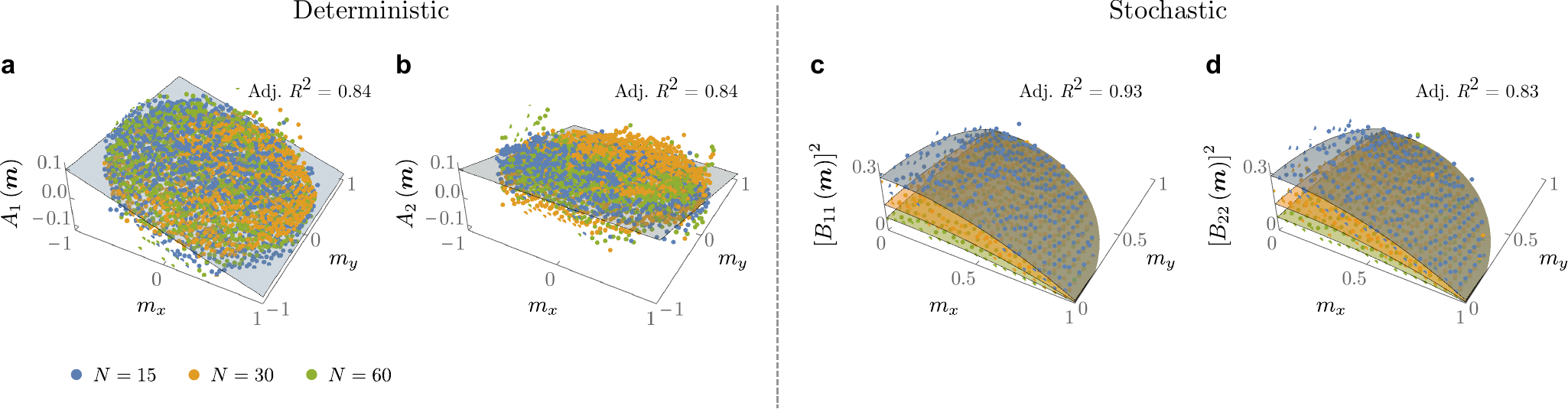}
	\caption
	{
		(Colour online) {\bf Dynamics: first and second jump-moments}.  The first jump-moments characterise the school's deterministic behaviour in the $N\to\infty$ limit (panels {\bf a} and {\bf b}).  Both $A_1$ and $A_2$ are characterised by a shallow linear slope with gradient $\approx-0.1$ and zero intercept, the former in the $m_x$ direction, and the latter in the $m_y$ direction. Dynamically, this corresponds to a weak pull towards isotropic, disordered motion ({\it i.e.}, $\boldsymbol{m}=\boldsymbol{0}$), irrespective of group size. The second jump-moments are related to the multiplicative pre-factors that determine the strength of the noise (panels {\bf d} and {\bf e}).  They are symmetric around, and maximal at, $\boldsymbol{m}=\boldsymbol{0}$, and zero for values $\vert\boldsymbol{m}\vert \gtrsim 1$.  The magnitude of the second jump moments, and hence the noise, are $O(1/\sqrt{N})$--- {\it i.e.}, they increase as group size decreases.  The cross-correlations $B_{12}$ and $B_{21}$ (not shown) are distributed randomly around zero with no trend.}
	\label{fig:3}
\end{figure*}
Consigning the details to Sec.~VI of the SM, we construct an It\^{o}-sense SDE from our data, which is of the general form
\begin{equation}
	\frac{d\boldsymbol{m}}{dt}=\boldsymbol{A}\left( \boldsymbol{m}\right) + \mathsf{B}\left( \boldsymbol{m}\right)\cdot\boldsymbol{\eta}(t),
	\label{eq:SDE}
\end{equation}
where $\boldsymbol{m}(t)\in \mathbb{R}^2$, bold typeface is used for two-dimensional vectors and sans-serif for a rank-2 tensor.  The elements of the vector $\boldsymbol{\eta}$ are independent sources of Gaussian white noise, such that $\langle\eta_j(t)\rangle = 0$ and $\langle\eta_j(t)\,\eta_k(t^\prime)\rangle = \delta_{jk}\,\delta(t-t^\prime)$ for $j,k=1,2$ (angle brackets indicates an average over stochastic realisations).

In principle, the components of the deterministic part of (\ref{eq:SDE}) can be found by numerically extracting the first jump moment(s) from the data.  Similarly, the components of the stochastic part can also be obtained by extracting the second jump moment(s), which are given by $\sum_k B_{jk}\,B_{lk}$ (where $B_{jk}$ are the components of $\mathsf{B}$).  In practice, however, this requires smooth interpolations, or `best fits' for the jump-moments.  Guided by exactly solvable one-dimensional toy models (see \cite{TBLDAJM14,Dyson2015} and Sec.~VIII, SM) we therefore propose and test different functional forms, each dependent on $N$, $\boldsymbol{m}$, and other unspecified parameters.  Using a simple least-squares procedure to fit the free parameters, we then choose expressions with the greatest adjusted-$R^2$, being careful to avoid over-fitting (see Fig.~\ref{fig:3}).

Substituting the resulting functions into (\ref{eq:SDE}), we obtain
\begin{equation}
	\frac{d\boldsymbol{m}}{dt}=-2\alpha\,\boldsymbol{m} + \left[\frac{2\beta\left(1- 
		\left\vert\boldsymbol{m}\right\vert^2\right) + 4\alpha}{N}\right]^{1/2}\,\mathsf{1}\cdot\boldsymbol{\eta}(t),
	\label{eq:dmdt}
\end{equation}
where $\mathsf{1}$ is the two-dimensional identity matrix, and the constants $\alpha=0.05$ and $\beta=2.0$ have been determined by the fitting procedure [integer pre-factors are chosen to highlight similarities with the aforementioned toy models (Sec.~VIII, SM)].  Reassuringly, direct Milstein-method numerical integration of (\ref{eq:dmdt}) recovers steady-state statistics that retain the key features of our experimental observations (Sec.~VII, SM).

Of note, we see that $\mathsf{B}$ is $O(1/\sqrt{N})$, whilst $\boldsymbol{A}$ is $O(\mathrm{constant})$, which confirms our earlier assumption and implies that the noise is likely {\it intrinsic}, due to probabilistic interactions between a finite number of individuals.  Moreover, in the deterministic $N\to\infty$ limit, the single stable fixed point of (\ref{eq:dmdt}) is at $\boldsymbol{m}=\boldsymbol{0}$, which corresponds to isotropic disordered motion.  This further implies that the observed high levels of ordering are in fact {\it noise-induced}, arising from the multiplicative noise term.

Informally, Eq.~(\ref{eq:dmdt}) can be understood in terms of the following heuristic; although the deterministic dynamics `pulls' the systems towards isotropic motion, the closer it gets, the larger the noise becomes, therefore `kicking' the system away, towards more aligned motion.  Conversely, the more aligned the motion, the smaller the fluctuations, and the longer the system is able to reside there.

\begin{figure*}[!t]
	\centering
	\includegraphics[width=0.98\textwidth]{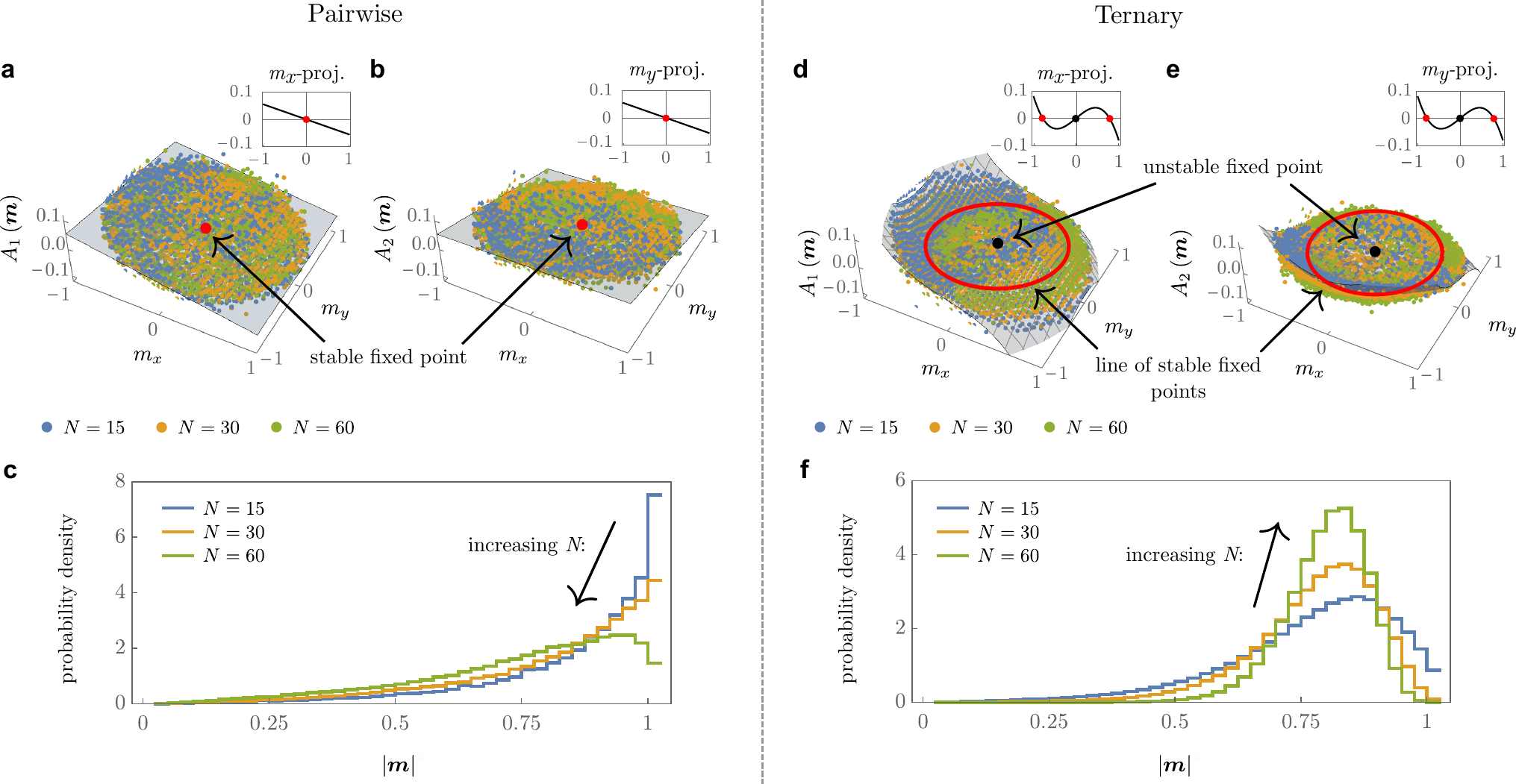}
	\caption
	{
		(Color online) {\bf Stochastic simulations}. Illustrative comparison between a pairwise-dominated model [{\it i.e.}, reactions (\ref{eq:random}) and (\ref{eq:copy}) with $s=0.25$ and $c=4$] and a ternary-dominated model [including the extra reaction defined by (\ref{eq:ternary_1}) and (\ref{eq:ternary_2}), where $s=0.25$, $c=0.01$ and $h=0.3$].  Whilst both approaches result in multiplicative noise that is similar to that observed in the experimental data (Sec.~VII, SM), only the pairwise model (panels {\bf a} and {\bf b}) reproduces the correct deterministic pull towards isotropic motion--- {\it i.e.}, $\boldsymbol{m}=\boldsymbol{0}$ ({\it cf.} Fig.~\ref{fig:3}, panels {\bf a} and {\bf b}).  For the ternary model, isotropic motion is an unstable fixed point, and a new line of (Lyapounov-)stable fixed points is introduced at finite $\vert\boldsymbol{m}\vert$ (panels {\bf d} and {\bf e}).  As a result, the $N$-dependence of the steady-state PDFs generated from the pairwise model closely resembles that derived from the data, whilst the ternary model does not (panels {\bf c} and {\bf f}).}
	\label{fig:4}
\end{figure*}
\section{A model of pairwise interactions reproduces salient features of the data}
Our analysis implies behaviour that is reminiscent of certain individual-based binary-choice models that appear in the literature \cite{Kirman1993,ALFARANO2007a,Altschuler2008,TBLDAJM14}, themselves loosely based on the voter model of opinion dynamics \cite{Liggett1999}.  By introducing continuous degrees of freedom, we propose a simple two-dimensional extension of such models, which makes notional contact with the mean-field XY model\footnote{However, the dynamical behaviour cannot be written in terms of the minimisation of a free-energy.}. 

We assume that at each instant in time, $t$, the direction of each fish 
\begin{equation}
	\hat{\boldsymbol{d}}_i(t) = \begin{pmatrix}
		\cos\theta_i(t)\\
		\sin\theta_i(t)
		\end{pmatrix},
	\label{eq:s}
\end{equation}
is, itself, prescribed by a stochastic protocol for the dynamics of the angles $\theta_i(t)$ ($i=1,\ldots,N$).  Using the notation of chemical reaction kinetics, we write down two competing types of underlying behaviour.

First, at a constant rate per unit time, $s$, every fish can spontaneously change its direction (angle).  That is
\begin{equation}
	\theta_i \xrightarrow{s} \theta_i + \mathcal{N}_\mathrm{trunc}\left( 0,\varepsilon, -\pi,\pi\right),
	\label{eq:random}
\end{equation}
where $\mathcal{N}_\mathrm{trunc}\left( \mu,\sigma^2,a,b \right)$ is a truncated normal distribution with mean $\mu$ and variance $\sigma^2$, normalised over the interval $(a,b)$.  

Second, at a different (but still constant) rate, $c$, a given fish `$i$' chooses another fish `$j$' and copies it--- {\it i.e.}, it turns to move in the same direction:
\begin{equation}
	\theta_i + \theta_{j\neq i} \xrightarrow{c} 2\theta_j.
	\label{eq:copy}
\end{equation}
This second equation describes a pairwise interaction, and assumes that the system is `mean-field' or `fully-connected'; a reasonable assumption given the small school sizes being considered.

Using the Gillespie algorithm \cite{Gillespie1976a,Gillespie1977}, we simulate the continuous-time Markov process represented by (\ref{eq:random}) and (\ref{eq:copy}), and compare our results with the above described experimental observations.  Specifically, for each school size ($N=15$, $30$, and $60$) we use simulations to generate steady-state probability distributions $P_{N,\,s,\,c,\,\varepsilon}\left(\left\vert \boldsymbol{m}\right\vert\right)$ and compare them to $P_N\left(\left\vert \boldsymbol{M}\right\vert\right)$, the distributions extracted from the experimental data--- {\it i.e.,} by normalising the histograms \ref{fig:2}{\bf g}, {\bf h} \& {\bf i}.  Scanning the  parameter space using a Genetic Algorithm \cite{Goldberg1989}, we find that, although each group size is slightly different, the values $s = 0.4$ , $c = 8$ and $\varepsilon = \pi$ broadly minimize the Kullback-Leibler (KL) divergence $D_{KL}\left[P_{N,\,s,\,c,\,\varepsilon}\left(\left\vert \boldsymbol{m}\right\vert\right)\vert\vert\,P_N\left(\vert\boldsymbol{M}\vert\right)\right]$, which is a measure of the information missing from the simulation-generated steady state PDF when compared with that of the real data.

The similarity between simulation-generated parameter values $s$ and $c$, and those extracted directly from the jump-moments, $\alpha$ and $\beta$, implies a more formal correspondence, especially in light of \cite{TBLDAJM14} (see Sec.~VIII, SM).  However, rigorously deriving an SDE from rules (\ref{eq:random}) and (\ref{eq:copy}) is not straightforward.  Nevertheless, we may extract the first and second jump-moments of the simulation-generated data numerically, as before.   Reassuringly, the results faithfully reproduce Eq.~(\ref{eq:dmdt}) and therefore all of the salient features of our experimental data (see Fig.~\ref{fig:4}, panels {\bf a} \& {\bf b}, and Fig.~4 of the SM).

\section{Higher-order interactions are sub-dominant}

Again drawing inspiration from simple models (see {\it e.g.}, \cite{Dyson2015,Jhawar2018} and Sec.~VIII, SM) we note that ternary or higher-order interactions cannot be dominant.  To understand this, consider the following ternary extension, to be included alongside interactions (\ref{eq:random}) and (\ref{eq:copy}):
\begin{equation}
\theta_i + \theta_{j\neq i} + \theta_{\substack{k\neq i\\ \ \neq j}} \xrightarrow{h}
\left\{\begin{array}{l}
2\theta_\ell + \theta_m\\
2\theta_m + \theta_\ell
\end{array}\right.,
\label{eq:ternary_1}
\end{equation}
where
\begin{equation}
\left\{\ell,m\right\} = 
\underset{\begin{subarray}{c} \left\{\{p,q\}\subset\{i,j,k\}\right\} \end{subarray}}{\mathrm{arg}\,\mathrm{min}}\left(\min\left(2\pi-\left\vert\theta_p-\theta_q\right\vert,\left\vert\theta_p-\theta_q\right\vert\right)\right).
\label{eq:ternary_2}
\end{equation}
That is, with a specific rate $h$, three individuals are picked at random and their directions are compared in order to find which pair have the minimum (acute) angular difference.  The remaining individual then turns to copy either of the two others with equal probability.

The resulting differences are clear (see Fig.~\ref{fig:4} and Sec.~VII, SM): ternary interactions effectively add a term $\sim h\,(1-\vert\boldsymbol{m}\vert)\,\boldsymbol{m}$ to the deterministic ($N\to\infty$) dynamics, significantly altering the $N$-dependence of the steady-state PDF.  In particular, if $h$ is sufficiently high, then the isotropic $\boldsymbol{m}=\boldsymbol{0}$ fixed point switches from being stable to unstable, and a line of new (Lyapounov-)stable fixed points appears at finite $\vert\boldsymbol{m}\vert$, meaning ordering is not lost as $N$ increases.  Instead, the steady-state probability distribution simply becomes increasingly peaked around the finite-$\vert\boldsymbol{m}\vert$ fixed point(s) associated with ternary interactions.

By contrast, if $h$ is sufficiently low that the fixed points of the deterministic dynamics remain unchanged then ternary interactions are sub-dominant, and the group-level behaviour essentially replicates that described for the pairwise only case.  That is, we cannot unambiguously rule out ternary, or indeed higher order interactions of any other type.  Nevertheless, the observed $N$-dependent ordering does appear to be clear evidence that the dominant mode of interaction is pairwise; involving only two fish, to the exclusion of all others (including local averaging).

The intuition provided by the above comparison is backed-up by more systematic simulations.  Specifically, we simulated generalisations of Eqs.~(\ref{eq:ternary_1},\ref{eq:ternary_2}) and a mean-field Vicsek model (the latter to explicitly rule out any effects arising from averaging), both of which included up to five-body interactions.  Then, using a Genetic Algorithm \cite{Goldberg1989} in the context of repeated Gillespie simulations, we optimised such many-body interaction models against the experimental data by using the KL-divergence.  In all cases, we found that the model most consistent with the data had a dominant mode of interaction that was pairwise--- {\it i.e.}, the rates corresponding to higher-order interactions were negligible (Sec.~VII, SM).

\section{Discussion}
In summary, we report rare empirical evidence of noise-induced schooling in fish due to finite-size effects--- serving to underline not only the importance of intrinsic noise, but also how it can manifest at the collective level as a multiplicative noise pre-factor.  In particular, our results have implications for the possible modes of interaction between fish, and what features these must have in order to give rise to such behaviour.

We put forward a generalisation of existing one-dimensional models which, although simple, convincingly reproduces not only the observed steady-state statistics, but also the nature of the jump-moments, and therefore the aforementioned noise-induced character.  It involves two types of behaviour: individuals can either spontaneously change direction, or copy another individual, chosen at random.  Importantly, this requires, at most, {\it pairwise} interactions, where a given fish only interacts with other fish, one at a time.  Simulations with dominant higher-order interactions, including local-averaging, do not represent the data well.  The reasons for this can be seen in the provided example of a ternary interaction\footnote{We stress that we have simulated up to five-body interactions.} whereby the deterministic dynamics is changed in a way reminiscent of the difference between quadratic and quartic potentials: a previously stable isotropic fixed point becomes unstable, and a line of new fixed points emerges at large polarisations, changing the dynamics dramatically.

Notably, these results are broadly in-line with Refs.~\cite{Herbert-Read2011,Katz2011a,Jiang2017} which concern very small groups of fish ($N \leq 5$), characterising correlations in turning angles, implied forces, and directions, respectively.  However, comparisons should also be drawn with \cite{Dyson2015} which, by contrast, employs a ternary-dominated model (albeit in one dimension) to describe the experimental observations of locust nymphs \cite{Yates2009a} mentioned earlier.  In this light, our work can be seen as evidence in support of a point argued-for in the introduction; the proper analysis of jump-moments can act as an important discriminating factor between different classes of behaviour and/or types of animal. 

We may also assess our results in the context of the {\it de facto} standard for modelling collective motion, the Vicsek model \cite{Vicsek1995}.  Along with other models in its class, the Vicsek model is based on individuals choosing their direction by performing an imperfect local average over their neighbours.  In a mean-field approximation, which ignores spatial degrees of freedom so that individuals are all mutual neighbours, we see that the so-called `Toner-Tu' hydrodynamic equations \cite{Toner1995} reduce to a single SDE for the group polarisation:
\begin{equation}
\frac{d\boldsymbol{m}}{dt} = -\left(a - b\,\vert\boldsymbol{m}\vert^2\right)\boldsymbol{m} + \boldsymbol{\eta},
\label{eq:viscek}
\end{equation}
which can be contrasted with Eq.~(\ref{eq:dmdt}). Firstly, the deterministic term is reminiscent of the cubic form found in both the ternary interaction model of \cite{Dyson2015} and that derived from our ternary interaction simulations [see Eq.~(10) of SM].  This implies that if the constant $b$ is too large, then higher order interactions will be dominant and poorly reflect the data.  Secondly, and perhaps most importantly, the noise is simply additive--- having been introduced {\it ad hoc}, rather than formally `derived'--- which leads to dramatically different dynamical behaviour (see Sec.~VIII of SM for further details).

In conclusion, our study unambiguously demonstrates the importance of noise due to probabilistic interactions between a finite number of individuals.  This suggests both a straightforward mechanism that might underpin group-size dependent effects (see {\it e.g.}, \cite{Gautrais2012a}), and more generally the need for a re-appraisal of traditional approaches to understanding collective motion.  Looking forwards, the dual issues of space and density are the clear challenges ahead. In this article, we have begun to understand how the total number of individuals, $N$, in localised schools, and therefore perhaps density in de-localised schools, non-trivially impacts on the ordering dynamics.  However, it is not clear {\it a priori} how the density within groups itself fluctuates, and to what extent those statistics are coupled to alignment. More generally, we are led to ask: how can the rigorous derivation of multiplicative noise terms be incorporated into broader active-hydrodynamic descriptions of collective motion \cite{Toner1995,Ramaswamy2010a}?  Some tentative theoretical steps have been made, notably in the context of both active nematics \cite{Bertin2013}, which extends previous work concerning (passive) Brownian particles \cite{DSD96}, and in one-dimensional models of direction-switching \cite{OLaighleis2018a,Chatterjee}.  We therefore welcome further work in the area.              

\section{Author Contributions}
 	VG conceived of, and oversaw the project.
 	JJ, UA-K and HR performed experiments.
 	JJ and RGM analysed and interpreted the data.
 	JJ and MDR performed simulations.
 	RGM wrote the manuscript, with input from JJ, MDR and VG.
	JJ and RGM contributed equally to the manuscript. 

\section{Declaration}
The authors declare no conflict of interest.

\section{Code Availablity}
Computer codes for image processing and data analysis can be found at: \texttt{https://github.com/tee-lab/schooling fish}. 

\section{Data Availability}
Sample (truncated) data files are included with the online code repository.  Full datasets will be made available for all reasonable requests. 

\section{Correspondence}
To whom correspondence should be addressed. E-mail: \texttt{guttal@iisc.ac.in}.

\section{Acknowledgements}
We acknowledge assistance from Subhakar Chakraborty and Elsa Mini Jos in the initial stages of experimental setup. We thank Binoy V V for suggestions on schooling fish species native to India and their hatcheries. We also thank S. Ramaswamy, for feedback / critical reading of the manuscript. JJ acknowledges support by the CSIR, India for research scholarship. RGM acknowledges both the Simons Foundation (USA) and EMBL-Australia for funding.  MDR acknowledges the Department of Science and Technology 'India-INSPIRE' award for funding. VG acknowledges support from DBT-IISc partnership program, infrastructure support from DST-FIST and SERB (DST).


\clearpage

\begin{widetext}

\section*{Supplementary Material}

\section{Introduction}\label{sec:intro}
This supplementary material comprises {\it i}) detailed descriptions of the experimental setup and image analysis, {\it ii}) details of theoretical calculations (including a terse summary of the necessary background and notation), and {\it iii}) additional data, which both supports and contextualises the core narrative of the main manuscript.  For sample data and analysis codes, we maintain a freely accessible online repository \cite{Github}.

\section{Experimental Details}\label{sec:setup}
{\it Etroplus suratensis} (local name `Karimeen’) is native to Sri Lanka and coastal parts of South India.  Based on anecdotal evidence of schooling by Etroplus in freshwater/estuarine aquatic ecosystems, we collected juveniles (length between 1.5 and 3 cm) from a hatchery located at Azhikode, Kerala, India (Regional shrimp hatchery, Azhikode, GPS location: 10.189197, 76.172245) and transported them to an experimental laboratory at IISc (Indian Institute of Science) Bengaluru, India. We kept all fish in a 250 litre capacity glass tank at a density of 100-120 fish per tank. Water was continuously filtered and oxygenated. Every fortnight, 50-60\% of water from the holding tank was replaced with fresh water. Fish were fed daily at 5 pm ($\pm$ 30 min). The holding tanks were exposed to ambient light and temperature; we estimate that the temperature of Bengaluru is roughly $3-6^\circ$C colder than their natural habitat. Despite this difference, qualitative observations confirmed that these fish readily schooled, both in the hatchery and in our laboratory. Fish were habituated to lab conditions at least for 15 days before conducting any experiments; the maximum wait period in the holding tank was 34 days. During this waiting period, any increase in the body size of fish was negligible.

Experiments were carried out in a white coloured rectangular tank (238 cm x 185 cm) made of fibreglass. To avoid clustering of individuals near the corners of the tank, a circular arena with a radius of 90 cm and a height of 10 cm was made using white sand. To ensure that the fish motion can be considered effectively two dimensional, the water level in the arena was kept at a height of 10 cm from the base of the tank, allowing the fish a very limited depth in which to swim. To record experiments, a camera (Canon EOS-600D) was mounted above the arena and controlled remotely from a computer away from the experiment. To minimize any vibrations from external sources, the tank was kept above polystyrene sheets of height 5 cm. The complete setup was covered by white opaque cloth to minimize external visual disturbances to fish and act as a light diffuser.

As explained in the main text, we investigated schooling for different group sizes. For each group size we carried out four identical trials. Water was filled at least one hour beforehand, with trials conducted during the day, between 11 am and 3 pm. The number of fish used for each trial--- either 15, 30, 60, 120 or 240--- was selected at random.  All fish were used for only one experiment. We transferred fish from the holding tanks to the experimental arena using a fishing net. The initial 20 minutes were considered an acclimatization period and hence no recordings were made during that time. After the acclimatization time, fish movements were video recorded using the overhead camera that was operated remotely. The duration of recording was $\sim 50$ mins at a spatial resolution of $1920 \times 1080$ pixels and temporal resolution of 25 frames per second--- {\it i.e.}, one frame every 0.04 s. For each size, four separate trials were conducted.

\begin{figure*}[!t]
	\centering
	\includegraphics[width=0.66\textwidth]{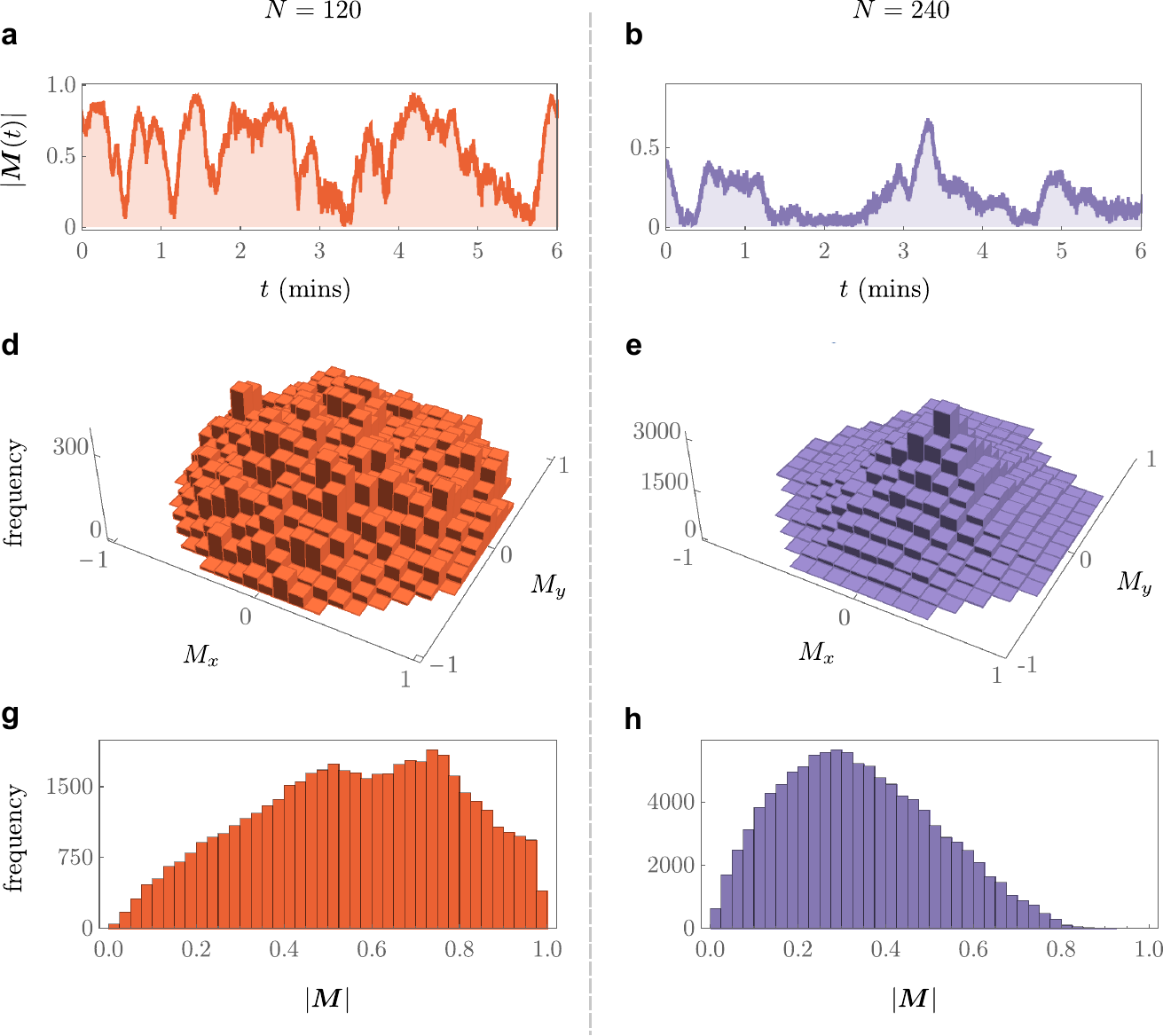}
	\caption
	{
		(Color online) {\bf Larger group sizes}. As with the smaller group sizes presented in the main manuscript, the trend continues for larger group sizes.  Panels d and e show that, for larger and larger $N$, the histogram on $\boldsymbol{M}$ becomes increasingly peaked around $\boldsymbol{M}=\boldsymbol{0}$ which, in turn, shifts mode further towards lower values in the corresponding histogram on $\vert\boldsymbol{M}\vert$ (panels g and h).  This is confirmed by visual inspection of individual stochastic trajectories (panels a and b).}
	\label{fig:S1}
\end{figure*}
We analysed the videos using a customized code based on packages available from the Image Processing Toolbox as part of the commercial software, Matlab \cite{Matlab}.  The tracking of individuals required three steps. The first step was to use image subtraction to identify the fish from their environment.  In the second step, we attributed an $(x,y)$-position to each fish by calculating the centroid of the obtained silhouettes.  Finally, using a Kalman filter approach \cite{Musoff2009} we tracked individuals, allowing us to obtain their velocities.

During one of the $N=60$ trials the fish remained (approximately) stationary for a long period of time at the start of the trial.  Whilst the inclusion of this data did not affect the overall conclusions of our work, it did skew the extracted parameter values (see following Sections and the main manuscript).  We therefore excluded the entire trial on the basis of insufficient acclimatisation period.    

The statistics of the velocities, and their combined polarisation ({\it cf.}~Eq.~(1) of the main manuscript) are time-independent.  For example, reconstructing Fig.~2 of the main manuscript, but using only 15 mins of data (as opposed to the entire 3.5 hour dataset) gives rise to slightly noisier, but nevertheless qualitatively indistinguishable results.  

\section{Larger Group Sizes}\label{sec:large}
In addition to the smaller groups discussed above and in the main text, we also explored two larger group sizes, $N=120$ and $N=240$ (see Fig.~\ref{fig:S1}). For such larger groups, we observe similar characteristics to the smaller group sizes.  For example, the mode of the histogram on $\vert\boldsymbol{M}\vert$ is shifted towards increasingly low values as $N$ increases.  However, observing the underlying videos of such large groups of fish, the spatial extent of the school can be quite large, often spanning a large portion of the tank. In particular, the local polarisation in one area of the tank, can occasionally be different to that of the entire group, implying that the mean-field analysis that we employ in the main manuscript is likely inappropriate. We therefore conservatively omit the results from our analysis, including them here for completeness only.

\begin{figure*}[!t]
	\centering
	\includegraphics[width=0.98\textwidth]{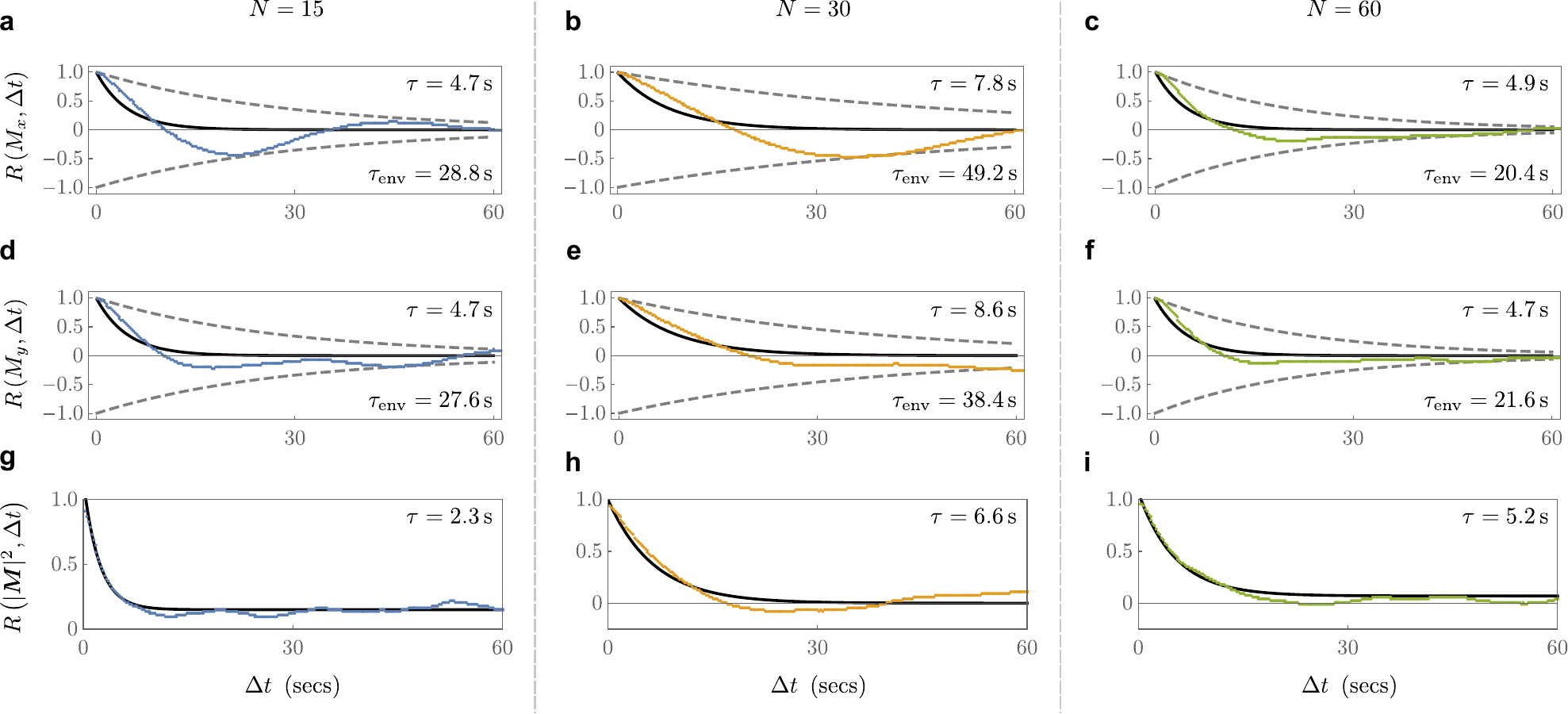}
	\caption
	{
		(Color online) {\bf Auto-correlations}. Blue, yellow and green solid lines represent the data for $N=15$, 30, and 60, resectively.  Two characteristic time-scales are apparent; $\tau$, which encapsulates the rate of initial decay of correlations to zero (solid black lines) and $\tau_\mathrm{env}$, which is rate of decay of the envelope of quasi-periodic correlations (dotted grey lines).}
	\label{fig:S2}
\end{figure*}
\section{Boundary Effects}\label{sec:bound}
A visual inspection of small-to-medium sized groups ($N=15$, $30$, and $60$) suggests limited interaction with the boundary.  More quantitatively, we follow \cite{Becco2006} and perform a control analysis by discarding all frames where any fish are within 15 cm, radially, from the tank wall ($\sim$ 6 fish body lengths). In doing so we lose a large portion of the data--- as much as $\sim80 \%$ for $N=30$--- but there is still sufficient information to derive an SDE.  

Reassuringly, the results remain consistent with those derived using the complete dataset: not only do steady-state histograms reproduce all the correct salient features but, fitting the first and second jump-moments reassuringly recovers the form of Eq.~(3) of the main manuscript, with coefficients $\alpha=0.05$ and $\beta=1.85$.  Adjusted-$R^2$ values are slightly lower than that of the full dataset, indicating an acceptable, but marginally worse fit.

We note that, if we remove not only the offending frame, but all subsequent frames within the correlation time $\tau\approx 6\,\mathrm{s}$ (see \S \ref{sec:corr}), we are not left with enough data to construct the requisite SDE with confidence.  For context therefore, we perform a random removal of the same number of frames as the `instantaneous' boundary control, in order to gauge the impact of the boundary.  Reassuringly, this gives rise to very similar results--- {\it i.e.}, qualitatively correct steady-state PDFs and an Eq.~(3) with coefficients $\alpha=0.05$ and $\beta=1.92$, whose quality of fit is, in fact, slightly better than the boundary-controlled data (see Table \ref{tab:boundary}).

If the boundary were significant to our findings, we would expect that the boundary control to provide a markedly worse fit than the random control, but this is not the case, and we therefore conclude that we are observing traits inherent to the behaviour of {\it E. suratensis}, rather than artefacts of our experimental setup.

\begin{table}
\caption{\label{tab:boundary}{\bf Controlling for boundary effects}. Removing frames in which fish are within 15 cms, radially, of the tank wall produces a fit comparable (indeed, marginally better) to that of removing the same number of frames, but at random.}
\begin{ruledtabular}
\begin{tabular}{lccc}
& \multicolumn{3}{c}{Adj.~$R^2$} \\

\cline{2-4}

& Full dataset & Boundary control & Random control \\
\hline

$A_1(\boldsymbol{m})$ & 0.84& 0.72& 0.68\\

$A_1(\boldsymbol{m})$ & 0.84& 0.72& 0.68\\

$\left[B_{11}(\boldsymbol{m})\right]^2$ & 0.93& 0.81& 0.77\\

$\left[B_{22}(\boldsymbol{m})\right]^2$ & 0.83& 0.69& 0.54\\

\end{tabular}
\end{ruledtabular}
\end{table}

\section{Autocorrelation Times}\label{sec:corr}
For each group size ($N=15$, $30$ and $60$) we calculated the auto-correlation of the time-indexed quantities $M_x$, $M_y$ and $\vert\boldsymbol{M}\vert^2$.  We used the generic form
\begin{equation}
R\left(X,\Delta t\right)=\frac{\left\langle \left(X_t-\mu\right)\left(X_{t+\Delta t} - \mu\right)\right\rangle}{\sigma^2},
\label{eq:R}
\end{equation}
where $X$ represents a set of $n$ time-indexed measurements with mean $\mu = \sum_t X_t/n$, and variance $\sigma = \sum_t \left(X_t-\mu\right)^2/n$.  The results are shown in Fig.~\ref{fig:S2}.

For the case of $M_x$ and $M_y$, the auto-correlation decays in an oscillatory manner (panels {\bf a}-{\bf f}).  We are therefore able to extract two characteristic times.  As argued-for in the main manuscript, the time-scale of behavioural interactions is approximated by fitting $\exp{\left(-\Delta t / \tau\right)}$ to the initial decay of the auto-correlation function, truncated where it first crosses zero.  We also fitted the envelope of the decay [$\exp{\left(-\Delta t / \tau_\mathrm{env}\right)}$] which we associate with the long-time correlations induced by finite tank size.

For the case of $\vert\boldsymbol{M}\vert^2$, the auto-correlation function is only weakly oscillatory, but appears to decay to a finite value as $\delta t$ becomes large.  We again attribute this to finite tank-size induced correlations, and find a characteristic decay time by fitting $\exp{\left(-\Delta t / \tau_{\vert\boldsymbol{M}\vert^2}\right)} + \mathcal{C}$, where both $\tau_{\vert\boldsymbol{M}\vert^2}$ and $\mathcal{C}$ are now fit-parameters. 

\section{Extracting Jump-Moments}\label{sec:extract}
With a formal basis in the system-size expansion of Master equations \cite{van_kampen,Gar03}, our basic assumption is that the data is well represented by an SDE of the form
\begin{equation}
	\frac{dm_i}{dt}=A_i\left(\boldsymbol{m}\right) + \sum_{j=1}^2 B_{ij}\left(\boldsymbol{m}\right)\,\eta_j,\ \forall\ i=1,2,
	\label{eq:SDE}
\end{equation}
which is just a component-wise version of Eq.~(2) in the main manuscript.  The coefficients $A_i$ and $B_{ij}$ are related to the first- and second-jump-moments--- $a_i^{(1)}$ and $a_{ij}^{(2)}$, respectively--- via:
\begin{equation}
	a_i^{(1)} = A_i\ \mathrm{and}\ a_{ij}^{(2)} = \sum_{k=1}^2 B_{ik}\,B_{jk},
	\label{eq:A&B}
\end{equation} 
and the $\eta_j$ are sources of delta-correlated Gaussian white noise with zero mean and unit variance [{\it i.e.}, $\left\langle \eta_j\right\rangle=0$, and $\left\langle \eta_i(t)\, \eta_j(t^\prime)\right\rangle = \delta_{ij}\,\delta(t-t^\prime)$].  More precisely, we assume that the data $\{\boldsymbol{M}(t_n)\in\mathbb{R}^2: \vert\boldsymbol{M}(t_n)\vert\leq1,\ t_n = n\,\delta t,\ \forall\ n=1,\ldots,\mathcal{N}\}$ satisfies a {\it discretised} version of (\ref{eq:SDE}).  Using the semi-formal notation prevalent within the physical sciences literature, we have
\begin{equation}
M_i(t_n + \Delta t) - M_i (t_n) = A_i\left(\boldsymbol{M}(t_n)\right)\,\Delta t + \sqrt{\Delta t}\sum_{j=1}^2 B_{ij}\left(\boldsymbol{M}(t_n)\right)\,\eta_j(t_n),
\label{eq:discrete}  
\end{equation}
where $\Delta t = \lambda\,\delta t$, for $\lambda \in\mathbb{Z}_+$ such that $1\leq\lambda\leq\mathcal{N}$.  At this stage, we may ask: what is the most appropriate value of $\lambda$ over which to cleanly extract jump-moments?  Dividing by $\Delta t$ and taking an average over the $\mathcal{N}$ independent instances of the noise (one for each data point in the time-series) gives
\begin{equation}
\left\langle
\frac{M_i(t_n + \Delta t) - M_i (t_n)}{\Delta t}
\right\rangle_\mathcal{N}
= A_i\left(\boldsymbol{M}(t_n)\right)+ \frac{1}{\sqrt{\Delta t}}\sum_{j=1}^2 B_{ij}\left(\boldsymbol{M}(t_n)\right)\,\left\langle\eta_j\right\rangle_\mathcal{N}.
\label{eq:A_i_1}
\end{equation}
Importantly, we note that, since $\mathcal{N}$ is finite, then the average indicated by angle brackets, $\langle\cdot\rangle_\mathcal{N}$, is only a {\it sample} mean.  That is, it is a stochastic variable itself, with a finite variance of $O(\mathcal{N})$.  However, we may also see that $\delta t = T/\mathcal{N}$--- {\it i.e.}, the smallest time-step in the system is just the total time period divided by the number of data points, $\mathcal{N}$.  Therefore $\Delta t = \lambda T / \mathcal{N}$ and so
\begin{equation}
\mathrm{Var}\left[
\left\langle
\frac{M_i(t_n + \Delta t) - M_i (t_n)}{\Delta t}
\right\rangle_\mathcal{N}\right]
\sim \frac{1}{\lambda\,T},
\label{eq:sample}
\end{equation}
which simply shows that the variance of the stochastic variable given by the right-hand side of (\ref{eq:A_i_1}) is just inversely proportional to $\lambda$, and hence the size of the time-step $\Delta t$.  Therefore, in order to obtain a good estimate for the function(s) $A_i$, the procedure is to use
\begin{equation}
	A_i\left( \boldsymbol{m}\right) = \frac{1}{\Delta t}\left\langle M_i\left( 
	t_n+\Delta t \right) - M_i\left( t_n 
	\right)\right\rangle_{\boldsymbol{M}(t_n)=\boldsymbol{m}},
	\label{eq:A_i}
\end{equation}
but choose a value of $\lambda$ that is large enough to minimise the noise arising from finite data whilst, at the same time, small enough to ensure $\Delta t$ is less than the correlation time, $\tau$ (see \S \ref{sec:corr}).  In the main manuscript, we use $\Delta t = \bar{\tau}$, the average decay time, across all trials, associated with the first zero of the autocorrelation function of polarisation components $M_x$ and $M_y$.

By contrast, extracting the coefficients $B_{ij}$ is relatively straightforward, and we may use the small size of $\Delta t$ to our advantage.  Multiplying two copies of Eq.~(\ref{eq:discrete}) together and dividing by $\Delta t$ gives
\begin{equation}
\frac{\left(M_i(t_n + \Delta t) - M_i (t_n)\right)\left(M_j(t_{n^\prime} + \Delta t) - M_j (t_{n^\prime})\right)}{\Delta t} = \sum_{k=1}^2 B_{ik}\left(\boldsymbol{M}(t_n)\right)\,\eta_k\,\sum_{\ell=1}^2 B_{i\ell}\left(\boldsymbol{M}(t_{n^\prime})\right)\,\eta_\ell + O\left(\sqrt{\Delta t}\right).
\end{equation}
Using the property that $\left\langle\eta_i(t)\eta_j(t^\prime)\right\rangle = \delta_{ij}\,\delta(t-t^\prime)$ then implies that
\begin{equation}
	\begin{split}
		\sum_{k=1}^2 B_{jk}\,B_{\ell k}\left( \boldsymbol{m}\right)
		&= \frac{1}{\Delta t}\left\langle 
		\left[M_j\left( t_n+\Delta t \right) - M_j\left( t_n 
		\right)\right]\right.\\&
		\quad\quad\quad\times\left.\left[M_\ell\left( t_n+\Delta t \right) - M_\ell\left( 
		t_n \right)\right]\right\rangle_{\boldsymbol{M}(t_n)=\boldsymbol{m}}.
	\end{split}
	\label{eq:B_ij}
\end{equation}
Here, in contrast to Eq.~(\ref{eq:A_i}), which requires the effect of the variance of the sample mean $\left\langle\eta_j\right\rangle$ to be countered, the optimal value of $\Delta t$ is just that which ensures the cleanest $O(\sqrt{\Delta t})$ cut-off, given the fixed sampling rate of the data.  This is given by $\lambda = 1$, and corresponds to setting $\Delta t = \delta t$, where $\delta t = 0.12$s.  

\begin{figure*}[!t]
	\centering
	\includegraphics[width=0.98\textwidth]{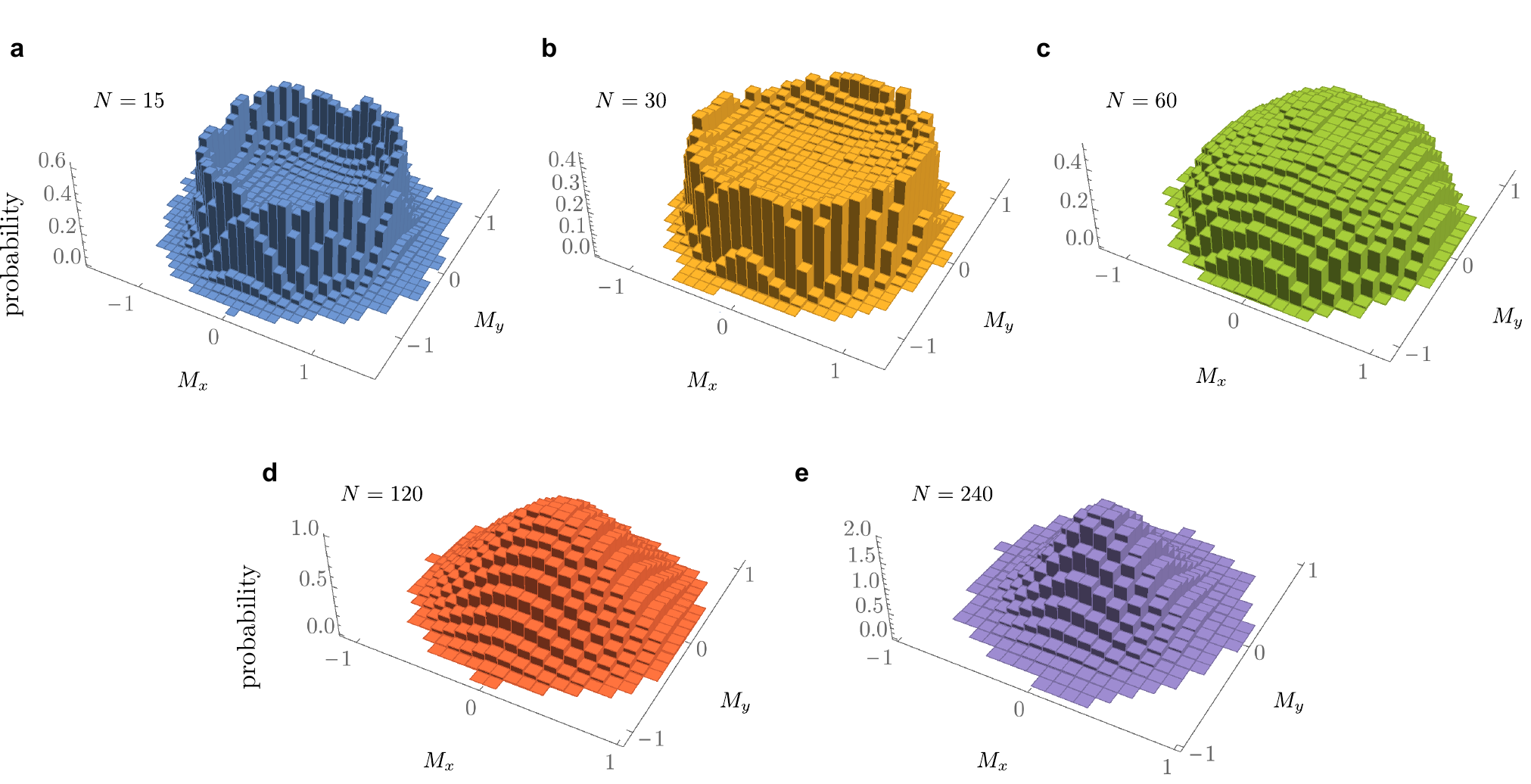}
	\caption
	{
		(Color online) {\bf Numerics}. Milstein-method simulations of the SDE that was extracted from the data via the fitting of jump-moments [Eq.~(5) of the main manuscript]. The results are qualitatively in-line with experimental observations, as expected.}
	\label{fig:S4}
\end{figure*}
Finally, we note that there are two practical considerations which must be taken into account.  First, not all values $\boldsymbol{M}$ are present in the dataset, and hence the averages on the right-hand sides of the above expressions should be taken with respect to $L$ small intervals: $\boldsymbol{m}_\Gamma \leq \boldsymbol{M}(t_n)\leq \boldsymbol{m}_\Gamma + \boldsymbol{\epsilon}$, where $\left\{\boldsymbol{m}_\Gamma = \Gamma \boldsymbol{\epsilon}: \vert\boldsymbol{\epsilon}\vert\ll 1,\,\Gamma=0,1,\dots,L\right\}$.  Secondly, the extracted jump-moments are likely noisy and it is typically necessary to find smooth interpolations, or `best fits', by proposing and testing different functional forms, each dependent on $N$, $\boldsymbol{m}$, and other unspecified parameters.

\section{Stochastic Simulations}

\subsection{Numerical Solutions of the Extracted SDE}
After extracting Eq.~(3) of the main manuscript from the data by fitting jump-moments, we used the Milstein-method implemented by the commercial software Mathematica \cite{wolf} to numerically generate solutions in order to ensure that it gives rise to the correct governing statistics (see Fig.~\ref{fig:S4}).

As a technical aside, we remark that these PDFs all suffer from the same deficiency.  They are defined on an open domain ($\mathbb{R}^2$) rather than the unit disk, which is the case for the experimental data.  The reason for this is that there is no systematic way to extract the necessary `reflecting-SDE' from either the data, or indeed a given Fokker-Planck equation.  To our knowledge, this issue arises in all such studies, including exactly solvable toy models, such as \cite{TBLDAJM14,Dyson2015}, which we discuss in \S \ref{sec:toy}.

\subsection{Jump Moments for the Ternary Model}
In Fig.~4 of the main manuscript we compare, for illustration purposes, Gillespie simulations of two proposed sets of `reactions'; one pairwise and the other ternary.  For brevity, only the coefficients of the first jump-moments are shown, since that is where the two models differ significantly.  For completeness, the fits for the second jump-moments are shown in Fig.~\ref{fig:S5}.  Here, both pairwise and ternary simulation data can be fitted by the functional form $\left[\left(2\beta^\prime + \gamma^\prime\right)\left(1-\vert\boldsymbol{m}\vert^2\right) + 4\alpha^\prime\right]/N$, which is inspired by analogous one-dimensional models that are solvable (see \S \ref{sec:toy}).  In each case, the fit parameters $\alpha^\prime$, $\beta^\prime$, and $\gamma^\prime$ take different values, reflecting the different simulation parameters $s$, $c$ and $h$.  Notably, in the pairwise case, when $h=0$, we have $\gamma^\prime=0$, which recovers the form used to fit the experimental data [{\it cf.}~Fig.~3 and Eq.~(3) of the main manuscript].  For the ternary model the corresponding SDE is omitted from the main manuscript for simplicity, and takes the form
\begin{equation}
	\frac{d\boldsymbol{m}}{dt} = -2\alpha^\prime\,\boldsymbol{m} + \frac{\gamma^\prime}{2}\left(1-\vert\boldsymbol{m}\vert^2\right)\boldsymbol{m} + \left[\frac{\left(2\beta^\prime + \gamma^\prime\right)\left(1-\vert\boldsymbol{m}\vert^2\right) + 4\alpha^\prime}{N}\right]^{1/2}\mathsf{1}\cdot\boldsymbol{\eta},
	\label{eq:tern_SDE}
\end{equation}
where we see explicitly that, although the multiplicative noise term remains very similar to the pairwise case, the deterministic $N\to\infty$ dynamics are clearly altered.  In particular, if $\gamma^\prime$ is sufficiently high, then the isotropic $\boldsymbol{m}=\boldsymbol{0}$ fixed point switches from being stable to unstable, and a line of new stable fixed points appears at finite $\vert\boldsymbol{m}\vert$.  As a result, ordering is not lost as $N$ increases--- a feature found in both pairwise simulations and experiments.  Instead, the steady-state probability distribution simply becomes increasingly peaked around the finite-$\vert\boldsymbol{m}\vert$ stable fixed point associated with ternary interactions.

\subsection{Higher-Order Copying Interactions}
\begin{figure*}[!t]
	\centering
	\includegraphics[width=0.98\textwidth]{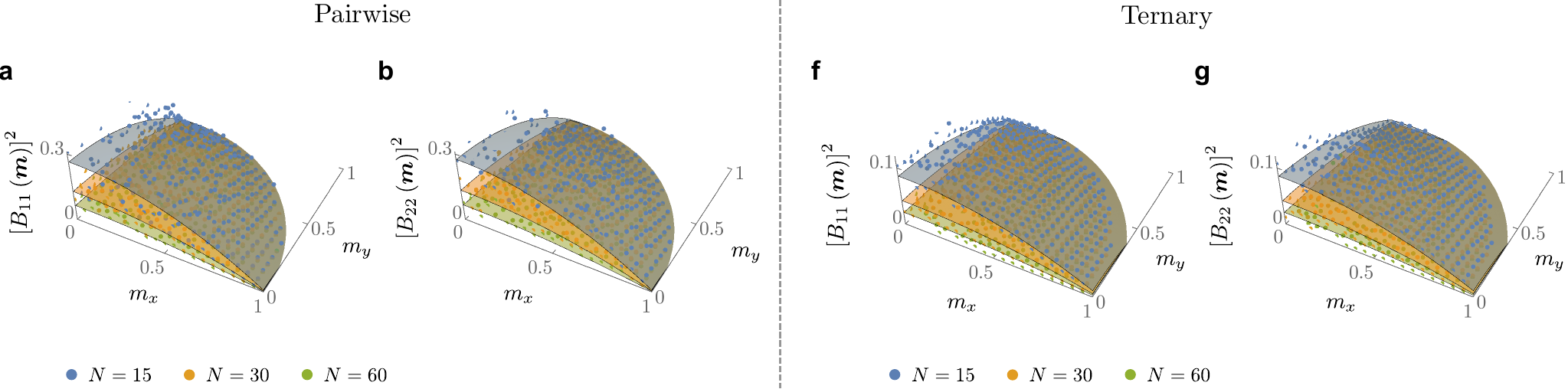}
	\caption
	{
		(Color online) {\bf Pairwise vs.~ternary: second jump-moments}. For both pairwise and ternary interaction models (see main manuscript) the coefficients of the second-jump moments can both be fitted by the generic form $\sim\left[\left(2\beta^\prime + \gamma^\prime\right)\left(1-\vert\boldsymbol{m}\vert^2\right) + 4\alpha^\prime\right]/N$, with different values for the fit parameters $\alpha^\prime$, $\beta^\prime$, and $\gamma^\prime$ (reflecting the different simulation parameters $s$, $c$ and $h$).  In both cases, this qualitatively resembles the jump moments extracted from the data
	}
	\label{fig:S5}
\end{figure*}
As mentioned above, the main manuscript describes an explicit example of a ternary copying interaction, which can be shown to poorly represent the data when the parameters take certain values.  More systematically, we considered a range of such individual-based $n$-body copying models, and used extensive computer simulations to optimise their parameters against the data.  Specifically, we considered mean-field $n$-body interactions that were of the form  
\begin{equation}
\theta_{i_1} + \theta_{i_2\neq i_1} + \ldots + \theta_{\substack{i_n\neq {i_1}\\ \ \ \neq i_2 \\\vdots\\ \\  \neq i_{n-1}}} \xrightarrow{r_n}
\left\{\begin{array}{l}
2\theta_{j_1} + \theta_{j_2}+\ldots+\theta_{j_{n-1}}\\
\theta_{j_1} + 2\theta_{j_2}+\ldots+\theta_{j_{n-1}}\\
\ \ \ \ \ \ \vdots \ \ \ \ \ \ \ \ \ \ \ \ \vdots \\
\theta_{j_1} + \theta_{j_2}+\ldots+2\theta_{j_{n-1}}\\
\end{array}\right.,
\label{eq:higher_order}
\end{equation}
where
\begin{equation}
\left\{j_1,j_2,\ldots,j_{n-1}\right\} = 
\underset{\begin{subarray}{c} \left\{ X \subset \left\{i_1,i_2,\ldots,i_n\right\}:\vert X\vert=n-1\right\} \end{subarray}}{\mathrm{arg}\,\mathrm{max}}\left\vert\sum_{k\in X} \begin{pmatrix}
		\cos\theta_k\\
		\sin\theta_k
		\end{pmatrix} \right\vert.
\label{eq:higher_order_2}
\end{equation}
This is a straightforward extension of the ternary interaction described in the main manuscript. With a specific rate $r_n$, $n$ individuals are picked at random and their directions are compared in order to find the $(n-1)$ individuals that are most aligned.  The remaining individual then turns to copy the direction of any of the others with equal probability.

In this context, we used the Gillespie algorithm \cite{Gillespie1976a,Gillespie1977} to simulate individuals subject to both spontaneous changes in direction [{\it cf.} Eq.~(5) of the main manuscript] and $n$-body copying interactions of the above form [Eqs.~(\ref{eq:higher_order}) and (\ref{eq:higher_order_2})].  We characterised each model by the integer $p$, such that copying interactions were included up to $p$-body.  That is, if $p=4$, the model contained 2-, 3-, and 4-body interactions {\it etc.}.  (Self-interactions have no meaning here).  As a result, each simulation required $p$ specific rates, $r_1,\ldots,r_p$ (with $r_1$ corresponding to spontaneous direction change).  Employing a Genetic Algorithm (GA) \cite{Goldberg1989}, Gillespie runs were then used generate steady-state PDFs $P_{N,\,r_1,\,\ldots,\,r_p,\,\varepsilon}\left(\left\vert \boldsymbol{m}\right\vert\right)$, which were optimised against the data, for a given $N$, by minimising the Kullback-Leibler (KL) divergence $D_{KL}\left[P_{N,\,r_1,\,\ldots,\,r_p,\,\varepsilon}\left(\left\vert \boldsymbol{m}\right\vert\right)\vert\vert\,P_N\left(\vert\boldsymbol{M}\vert\right)\right]$.  Once the optimal rates had been found, we re-computed the KL divergence many times in order to obtain statistics that further account for the inherent variation associated with different Gillespie runs (using the same parameters).

The results are shown in Table \ref{tab:GA_Copy}, where it is clear that, in each instance, the GA returns models whose fit with the steady-state statistics of the data is very good, and whose specific rates are negligible for all $n$- body interactions above pairwise ({\it i.e.}, $n>2$).  That is, the data (and its underlying noise-induced character) can be viewed as strong evidence underlying interactions that are pairwise-dominant.  

As a technical aside, we note that, in practice, the KL divergence differs slightly with each Gillespie `run'.  However, repeated runs using the GA-optimised rates indicate that the small differences in KL divergence that appear in Table \ref{tab:GA_Copy} for different models--- {\it i.e.,} different $p$--- are well within this expected statistical variation.   

\begin{table}
\caption{\label{tab:GA_Copy}{\bf Optimisation of higher-order copying interaction models}.  Using a Genetic Algorithm in the context of repeated Gillespie simulations, we optimise a given model's specific rates against the experimental data.  The results imply that pairwise copying is the dominant mode of interaction, and higher order interactions are likely negligible.}
\begin{ruledtabular}
\begin{tabular}{lllcc}

$N$& $p$ & Rates & $\overline{D_{KL}\left[P_{N,\,r_1,\,\ldots,\,r_p,\,\varepsilon}\left(\left\vert \boldsymbol{m}\right\vert\right)\vert\vert\,P_N\left(\vert\boldsymbol{M}\vert\right)\right]}$
& $\sigma^2\left\{D_{KL}\left[P_{N,\,r_1,\,\ldots,\,r_p,\,\varepsilon}\left(\left\vert \boldsymbol{m}\right\vert\right)\vert\vert\,P_N\left(\vert\boldsymbol{M}\vert\right)\right]\right\}$\\
\hline
\multirow{14}{*}{15}& 2 & $r_1=0.48$ & 0.0504& 0.0012\\
& &$r_2=7.99$ & & \\
\cline{2-5}
&3&$r_1=0.47$ & 0.0503& 0.0012\\
& &$r_2=7.40$ & & \\
& &$r_3=0.01$ & & \\
\cline{2-5}
&4&$r_1=0.48$ & 0.0498& 0.0013\\
& &$r_2=7.69$ & & \\
& &$r_3=0.00$ & & \\
& &$r_4=0.00$ & & \\
\cline{2-5}
&5&$r_1=0.48$ & 0.0500& 0.0012\\
& &$r_2=7.19$ & & \\
& &$r_3=0.00$ & & \\
& &$r_4=0.00$ & & \\
& &$r_5=0.00$ & & \\
\hline
\multirow{14}{*}{30}& 2 & $r_1=0.32$ & 0.0314& 0.0018\\
& &$r_2=8.41$ & & \\
\cline{2-5}
&3&$r_1=0.32$ & 0.0312& 0.0016\\
& &$r_2=8.59$ & & \\
& &$r_3=0.00$ & & \\
\cline{2-5}
&4&$r_1=0.35$ & 0.0322& 0.0016\\
& &$r_2=8.46$ & & \\
& &$r_3=0.04$ & & \\
& &$r_4=0.02$ & & \\
\cline{2-5}
&5&$r_1=0.33$ & 0.0320& 0.0017\\
& &$r_2=8.41$ & & \\
& &$r_3=0.01$ & & \\
& &$r_4=0.00$ & & \\
& &$r_5=0.02$ & & \\
\hline
\multirow{14}{*}{60}& 2 & $r_1=0.23$ &0.0460 &0.0042 \\
& &$r_2=8.49$ & & \\
\cline{2-5}
&3&$r_1=0.24$ &0.0478 & 0.0044\\
& &$r_2=8.41$ & & \\
& &$r_3=0.01$ & & \\
\cline{2-5}
&4&$r_1=0.24$ & 0.0475& 0.0039\\
& &$r_2=8.65$ & & \\
& &$r_3=0.00$ & & \\
& &$r_4=0.01$ & & \\
\cline{2-5}
&5&$r_1=0.25$ & 0.0478& 0.0047\\
& &$r_2=8.72$ & & \\
& &$r_3=0.02$ & & \\
& &$r_4=0.01$ & & \\
& &$r_5=0.00$ & & \\
\end{tabular}
\end{ruledtabular}
\end{table}

\subsection{Higher-Order Vicsek-like Interactions}
Given its prevalence in the collective motion literature, we wanted to explicitly compare our results with any Vicsek-like behaviour, where alignment results from individuals averaging the directions of a number of neighbours.  For comparison with our existing results, we consider an individual-based analogue of the Vicsek model that is mean-field--- {\it i.e.}, it does not describe spatially distributed collectives, and all individuals are technically `neighbours'.  Here, individuals can perform one of two actions.  Specifically, they can change their direction at random [{\it cf}.~Eq.~(5) of the main manuscript], or they choose $n-1\leq N$ other individuals, and turn to move towards the average direction of those individuals .  That is
\begin{equation}
	\theta_{i_1} + \theta_{i_2\neq i_1} + \ldots + \theta_{\substack{i_n\neq {i_1}\\ \ \ \neq i_2 \\\vdots\\ \\ \ \ \neq i_n}} \xrightarrow{r_n}
	\frac{1}{n-1}\sum_{j=2}^n \theta_{i_j} + \theta_{i_2}+\ldots+\theta_{i_n}.
	\label{eq:higher_order_Vicsek}
\end{equation}
This has the benefit that, for $n=2$, we recover the pairwise copying interaction used throughout our study, and yet higher order terms follow the canonical direction-averaging protocol of the Vicsek model in the limit of no error.

Once again using a GA to scan the relevant parameter space, repeated Gillespie simulations indicate that the KL-divergence between the PDF extracted from the data and that generated from simulations is minimised by pairwise interactions--- {\it i.e.}, $n=2$--- and that any kind of higher-order direction averaging results in a significant mismatch that cannot be attributed to the inherent fluctuations associated with Gillespie simulations (see Table \ref{tab:GA_Vicsek}).

\begin{table}
\caption{\label{tab:GA_Vicsek}{\bf Optimisation of higher-order Vicsek-like interaction models}.  Using a Genetic Algorithm in the context of repeated Gillespie simulations, we optimise a given model's specific rates against the experimental data.  The results confirm that direction-averaging is not represented by the data, and that pairwise copying interactions are the dominant behaviour.}
\begin{ruledtabular}
\begin{tabular}{lllcc}

$N$& $n$ & Rates & $\overline{D_{KL}\left[P_{N,\,r_1,\,r_n,\,\varepsilon}\left(\left\vert \boldsymbol{m}\right\vert\right)\vert\vert\,P_N\left(\vert\boldsymbol{M}\vert\right)\right]}$
& $\sigma^2\left\{D_{KL}\left[P_{N,\,r_1,\,r_n,\,\varepsilon}\left(\left\vert \boldsymbol{m}\right\vert\right)\vert\vert\,P_N\left(\vert\boldsymbol{M}\vert\right)\right]\right\}$\\
\hline
\multirow{8}{*}{15}& 2 & $r_1=0.48$ & 0.0504& 0.0014\\
& &$r_2=7.99$ & & \\
\cline{2-5}
&3&$r_1=0.85$ & 0.2250& 0.0022\\
& &$r_3=7.16$ & & \\
\cline{2-5}
&4&$r_1=1.07$ & 0.3228& 0.0024\\
& &$r_4=6.28$ & & \\
\cline{2-5}
&5&$r_1=1.22$ & 0.3821& 0.0018\\
& &$r_2=6.87$ & & \\
\hline
\multirow{8}{*}{30}& 2 & $r_1=0.32$ & 0.0314& 0.0018\\
& &$r_2=8.41$ & & \\
\cline{2-5}
&3&$r_1=1.45$ & 0.4062& 0.0047\\
& &$r_3=9.20$ & & \\
\cline{2-5}
&4&$r_1=1.95$ & 0.5967& 0.0042\\
& &$r_4=7.21$ & & \\
\cline{2-5}
&5&$r_1=2.47$ & 0.7179& 0.0045\\
& &$r_5=8.44$ & & \\
\hline
\multirow{8}{*}{60}& 2 & $r_1=0.23$ &0.0460 &0.0042 \\
& &$r_2=8.49$ & & \\
\cline{2-5}
&3&$r_1=2.31$ &0.5710 & 0.0068\\
& &$r_3=8.36$ & & \\
\cline{2-5}
&4&$r_1=3.94$ & 0.7408& 0.0065\\
& &$r_4=8.71$ & & \\
\cline{2-5}
&5&$r_1=5.58$ & 0.8545& 0.0065\\
& &$r_2=9.29$ & & \\
\end{tabular}
\end{ruledtabular}
\end{table}

\section{Supplementary Material}

\section{One-dimensional Toy Models}\label{sec:toy}

As described in the main text, our findings are reminiscent of a series of one-dimensional models which, whilst too simplified to explain fish motion in two-dimensions, are still helpful since they retain certain core characteristics, particularly in the context of understanding the differences between pairwise and ternary interactions.  The models in question involve a `binary choice' between two outcomes, typically assumed to be directions of motion along a line.  They take the form of mesoscopic SDEs which, due to their simplified nature, can be derived directly from microscopic rules very similar to those proposed in the main manuscript (rather than extracted from computer simulations, as we have done).  In all cases, the system is fully described by a single variable $x\in\left[-1,1\right]$, which is the concentration of individuals moving in either direction along the line--- {\it e.g.}, if $x=1$, all individuals are moving to the right, whilst if $x=-1$, all individuals are moving to left {\it etc}.  The SDEs therefore take the general form
\begin{equation}
	\frac{dx}{dt}=f(x) + g(x)\,\eta(t),
	\label{eq:1DSDE}
\end{equation}
where $\eta$ is delta-correlated Gaussian white noise with zero mean and unit variance.  The finer details of such models have been published elsewhere, and have even been contrasted as a pedagogical exercise in \cite{Jhawar2018}.  We therefore highlight only the points salient to the arguments made here, referring the reader to the relevant references where possible.

The 1-dimensional analogue of our pairwise interaction model has been introduced in many contexts, including foraging animals \cite{Kirman1993,TBLDAJM14}, financial trading \cite{ALFARANO2007a} and even cell signalling \cite{Altschuler2008}.  We follow closely the approach presented in \cite{TBLDAJM14}, which concerns foraging ants.  As echoed by the model proposed in the main text, the ants have two types of behaviour; at given specific rates $a$ and $b$, respectively, they either switch direction or choose another ant and copy its direction.  This results in an SDE that takes the form
\begin{equation}
	\frac{dx}{dt} = -2\,a\,x + \left[\frac{2b\left(1-x^2\right) + 4a}{N}\right]^{1/2}\eta(t),
	\label{eq:ants}
\end{equation}
and whose characteristics are plotted in Fig.~\ref{fig:S5}, panels {\bf b}, {\bf e} and {\bf h}.  Here we see a simple deterministic dynamics which is equivalent to motion in a quadratic potential, and has a single stable fixed point at $x=0$, which corresponds to isotropic motion.  Conversely, the multiplicative noise pre-factor is largest at $x=0$, decreasing to zero as $x$ approaches $\pm 1$.  It is also dependent on the system-size and has an overall scaling proportional to $1/\sqrt{N}$.  The resulting behaviour is captured by the steady state PDF, which can be solved for, and takes the form
\begin{equation}
	P_s(x) = \frac{1}{\mathcal{P}_0} \left[4a + 2b(1-x^2)\right]^{\frac{a\,N}{b}-1},
	\label{eq:ss_ant}
\end{equation}
where $\mathcal{P}_0$ is a normalisation factor.  As can be seen from Fig.~\ref{fig:S5}{\bf h}, this expression demonstrates a clear noise-induced behaviour whereby, for small values of $N$, the noise is large and the most likely states are $x=\pm 1$, which corresponds to coherent, polarised motion.  By contrast, for larger $N$, the mode of the distribution shifts to the deterministic fixed point at $x=0$--- {\it i.e}., isotropic motion--- with the distribution becoming more tightly peaked as $N$ increases further.

\begin{figure*}[!t]
	\centering
	\includegraphics[width=0.98\textwidth]{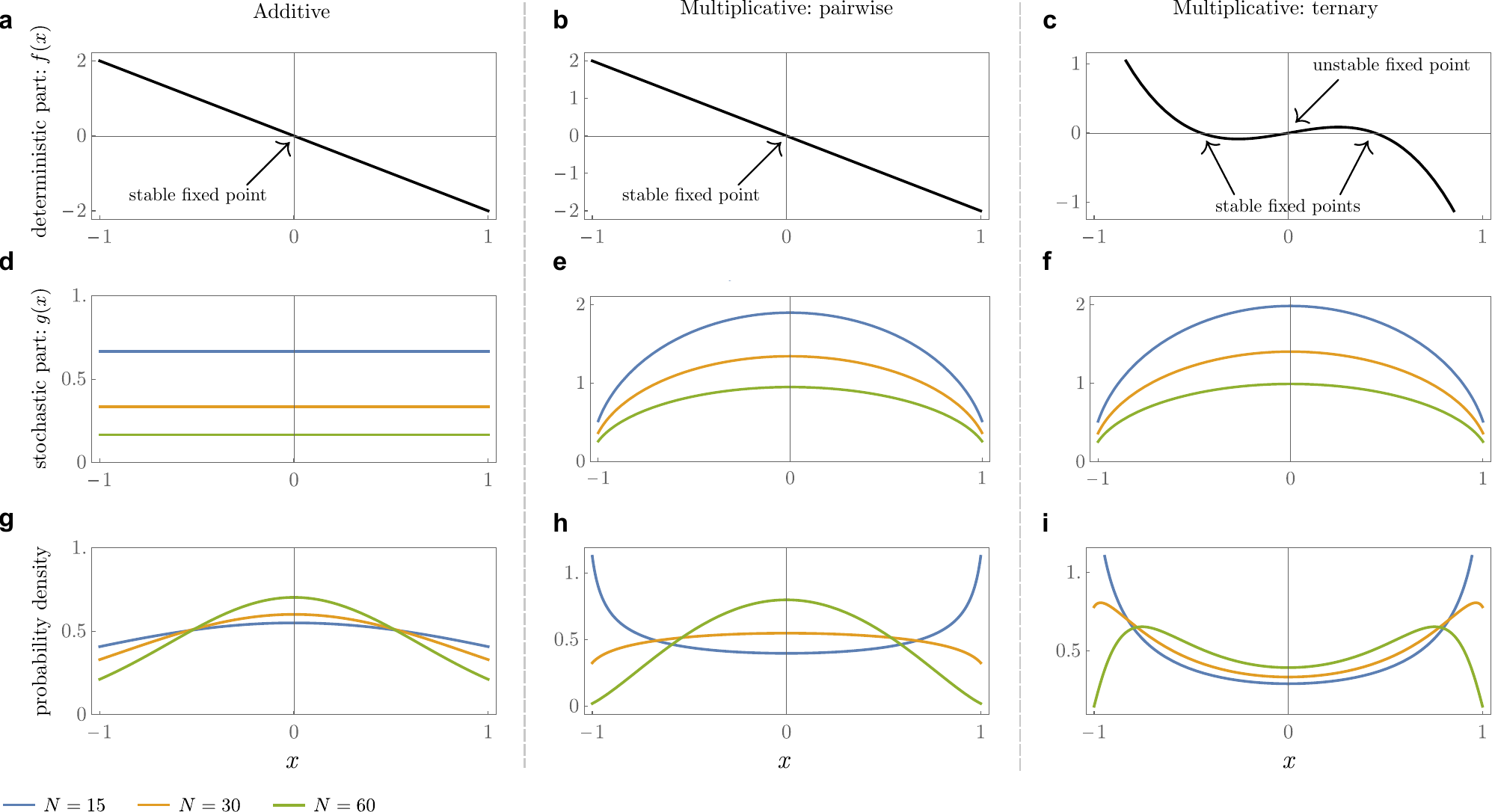}
	\caption
	{
		(Color online) {\bf Solvable toy-models}. Our results are indicative of behaviour that can be demonstrated in exactly solvable one-dimensional models.  Panels a, d \& g correspond to a simple attractive potential with additive noise, whose magnitude is $N$-dependent [see Eq.~\ref{eq:additive}].  Panels b, e \& h correspond to the one-dimensional equivalent \cite{TBLDAJM14} to the pairwise model proposed in the main manuscript [also Eq.~(\ref{eq:ants})].  Panels c, f \& i correspond to the one-dimensional equivalent \cite{Dyson2015} of the ternary model proposed in the main manuscript [also Eq.~(\ref{eq:locust})].  The values $a=1$, $b=25$, $c=5$, and $\sigma=10$ were used to plot these figures.}
	\label{fig:S6}
\end{figure*}
The effects of ternary interactions can be examined by adding the following reaction, used in \cite{Dyson2015} in the context of locusts marching around a ring.  At a rate $c$, three individuals are picked and the individual who is moving in the opposite direction to the other two turns to follow them.  This changes the form of the governing SDE, which becomes
\begin{equation}
	\frac{dx}{dt} = -2\,a\,x + \frac{c\,x}{2}\left(1-x^2\right) + \left[\frac{\left(2b+c\right)\left(1-x^2\right) + 4a}{N}\right]^{1/2}\eta(t).
\label{eq:locust}
\end{equation}   
The main difference with (\ref{eq:ants}) is that the deterministic part is now cubic, such that the previously stable $x=0$ fixed point is now unstable, and two new stable fixed points appear at $\pm x^\ast$, where $0<\vert x^\ast\vert <1$.  The stochastic part is qualitatively similar to the pairwise case.  Nevertheless, the steady-state PDF is markedly different, given by
\begin{equation}
	P_s(x) = \frac{1}{\mathcal{P}_0} \left[4a + (2b+c)(1-x^2)\right]^{\frac{4\,N\,a\,(b+c)}{(2b + c)^2}-1}\,\mathrm{exp}\left[\frac{c\,x^2\,N}{2(2b +c)}\right].
\label{eq:ss_locust}
\end{equation}
Plotting this expression, it is clear that, as before, when $N$ is small, the multiplicative noise ensures a noise-induced state, with modes at $x=\pm 1$.  However, for larger values of $N$, the system does not peak around the isotropic fixed point at $x=0$, since it is unstable.  Instead, we see symmetric peaks begin to form at intermediate values of $\vert x\vert$ which reflects an interplay between the multiplicative noise and the new stable fixed points associated with the ternary interaction.

In order to draw comparisons with the mean-field Vicsek model in 1-dimension, we contrast both of these cases with a simple 'null' model, which is formulated not by writing-down microscopic rules and systematically performing a system-size expansion, but instead by writing-down the mean-field dynamics associated with the given microscopics, and then simply adding noise in an {\it ad-hoc} fashion.  For the case where individuals randomly change their direction, left or right, with a specific rate $a$, the aforementioned method leads to an equation of the form
\begin{equation}
	\frac{dx}{dt} = -2\,a\,x + \frac{\sigma}{\sqrt{N}}\,\eta(t),
	\label{eq:additive}
\end{equation}
which corresponds to the following steady state PDF
\begin{equation}
	P_s(x)=\frac{1}{\mathcal{P}_0}\mathrm{Exp}\left[\frac{-2\,N\,\alpha\,x^2}{\sigma^2}\right].
	\label{eq:ss_additive}
\end{equation}
These expressions are equivalent to those of a particle moving in a quadratic potential that is subject to an $N$-dependent 'temperature'.  As a result, the larger the system size, the more tightly peaked the PDF around the deterministic fixed point at $x=0$.  This is, of course, fundamentally different to the multiplicative noise that arises from a systematic expansion of the master equation, and the type of behaviour that is seen in stochastic simulations, such as the Gillespie algotrithm.

\end{widetext}

\end{document}